\documentclass[lettersize,journal]{IEEEtran}
\usepackage{amsmath,amsfonts}
\usepackage{algorithmic}
\usepackage{array}
\usepackage{textcomp}
\usepackage{stfloats}
\usepackage{url}
\usepackage{verbatim}
\usepackage{graphicx}
\usepackage{cite}
\usepackage{multirow}
\usepackage{subfigure}
\usepackage{makecell}
\usepackage{diagbox}
\usepackage{color}
\def\BibTeX{{\rm B\kern-.05em{\sc i\kern-.025em b}\kern-.08em
    T\kern-.1667em\lower.7ex\hbox{E}\kern-.125emX}}
\usepackage{balance}
\begin{document}
\title{SNN-SC: A Spiking Semantic Communication Framework for Collaborative Intelligence}
\author{Mengyang Wang, Jiahui Li, \IEEEmembership{Member, IEEE}, Mengyao Ma, \IEEEmembership{Member, IEEE}, \\ Xiaopeng Fan, \IEEEmembership{Senior Member, IEEE}

\thanks{This work was supported in part by the National Key R\&D Program of China (2021YFF0900500), the National Natural Science Foundation of China (NSFC) under grants U22B2035 and 62272128.  \textit{(Corresponding author: Xiaopeng Fan.)}

Mengyang Wang and Xiaopeng Fan are with the School of Computer Science, Harbin Institute of Technology, Harbin 150001, China, and also with Peng Cheng Laboratory, Shenzhen 519055, China (e-mail: mywang1996@outlook.com; fxp@hit.edu.cn). 

Jiahui Li and Mengyao Ma are with the Wireless Technology Lab, Huawei, Shenzhen 518129, China (e-mail: lijiahui666@huawei.com; ma.mengyao@huawei.com). 
}
}

\markboth{Journal of \LaTeX\ Class Files,~Vol.~18, No.~9, September~2020}%
{How to Use the IEEEtran \LaTeX \ Templates}

\maketitle

\begin{abstract}
Collaborative Intelligence (CI) has emerged as a promising framework for deploying Artificial Intelligence (AI) models on resource-constrained edge devices. In CI, the AI model is partitioned between the edge device and the cloud, with intermediate features transmitted from the edge sub-model to the cloud sub-model to complete the inference task. However, reducing feature transmission overhead while maintaining task performance remains a challenge, particularly in the case of noisy wireless channels. In this paper, we propose a Spiking Neural Network (SNN)-based Semantic Communication (SC) model, SNN-SC, which extracts compact semantic information from features and transmits it through digital binary channels. Compared to the Deep Neural Network (DNN)-based SC model, whose output is floating-point, the binary output of SNN makes SNN-SC directly applicable to digital binary channels without the need for extra quantization. Moreover, we introduce a novel spiking neuron called IHF to enhance the reconstruction capability of the SNN-SC decoder. Finally, we enhance the performance of SNN-SC by maximizing the entropy of semantic information. SNN-SC achieves a higher compression ratio and overcomes the `cliff effect' compared to the traditional separate source and channel coding method. In addition, SNN-SC has lower computational complexity than the DNN-based SC model and maintains higher task performance under poor channel conditions.

\end{abstract}

\IEEEpeerreviewmaketitle

\section{Introduction}

\IEEEPARstart{I}{n} autonomous driving, vehicles require the assistance of artificial intelligence models to analyze the collected data. However, due to the limited computing resources of the hardware on vehicles, directly executing complex models is often impractical. To solve this problem, recent studies have proposed a framework based on edge-cloud collaborative reasoning, called Collaborative Intelligence (CI) \cite{2}. CI splits the complex Deep Neural Network (DNN) model into two parts: the edge device part and the cloud part, which collaborate to complete the computation of the entire model \cite{5, 69}. Edge devices like vehicles process the collected data by parts of their models to extract features. These features are then transmitted to the cloud server for subsequent inference tasks \cite{7, 64}. However, transmitting the original feature requires a large amount of channel bandwidth, and the channel between the device and the cloud is imperfect. Thus, it is critical to encode the feature into a compact and robust representation, which can be achieved by Semantic Communication (SC).

In contrast to conventional communication, SC focuses on the successful transmission of semantic information conveyed by the source rather than the accurate reception of every single symbol or bit \cite{53}. In other words, SC only transmits necessary information relevant to the specific task at the receiver, resulting in a significant reduction in bandwidth consumption \cite{85}. With the emergence of Deep Learning (DL) \cite{1}, several DL-enabled SC models have shown great potential for transmitting images \cite{17, 18, 19, 20}, texts \cite{54, 55, 56}, and deep features \cite{21, 22, 23, 24}. To enhance the robustness of the model, most SC models take noisy channels as non-trainable layers during training. Notably, the SC model for feature transmission is an effective tool in CI \cite{4}, which reduces feature transmission overhead while maintaining task performance in the cloud.

Nevertheless, due to the floating-point output of DNN, most SC models focus on transmitting features over analog channels, such as Additive White Gaussian Noise (AWGN) channel. In practice, digital channels like Binary Symmetric Channel (BSC) and Binary Erasure Channel (BEC) are also widely used. To ensure compatibility with digital channels, existing studies \cite{22,92,93} introduced quantization or sampling steps to binarize the information before transmission, facing the following challenges. Firstly, extra quantization or sampling unavoidably increases computational overhead. Secondly, errors in high-weight bits of the quantized information can significantly degrade task performance, especially under poor channel conditions. To meet these challenges, we creatively introduce Spiking Neural Network (SNN) to construct the feature SC model.


SNN is known as the third generation of neural network model \cite{25}, which models the behavior of biological neurons \cite{26}. Compared with DNN, SNN has the following advantages. Firstly, SNN employs discrete spike signals instead of analog signals to transmit information, which renders it more amenable to hardware implementation \cite{59}. Secondly, the binary spike signals in SNN convert the traditional high-power Multiply-Accumulation (MAC) into low-power Accumulation (AC), thereby resulting in a significant reduction in computational complexity and energy consumption \cite{32,47, 79}. Thirdly, the spiking neuron accumulates the input as its membrane potential at each time step, firing a spike only when the membrane potential exceeds a threshold, which equips SNN with inherent denoising capabilities\cite{28,29}.  These characteristics of SNN make it efficiently compatible with digital channels. 

In this paper, we propose a new SC framework based on SNN, SNN-SC, to transmit features over digital channels. SNN-SC not only eliminates the need for explicit quantization but also has powerful noise resiliency capability. SNN-SC is trained in a task-oriented manner, enabling it to extract semantic information from features efficiently. To evaluate the effectiveness of SNN-SC, we compare it with existing SC models and traditional communication model. Existing SC models are structurally similar to SNN-SC, but use quantization or sampling to binarize the semantic information. Traditional model uses source coding and channel coding to compress and transmit features. To the best of our knowledge, this is the first work to build an SNN-based feature SC model. The main contributions are summarized as follows:

\begin{enumerate}[\topsep=0pt]
\item We propose an SNN-based feature SC framework for CI under digital channels, named SNN-SC, where the transceiver is jointly optimized to perform efficient compression and robust transmission of features. With the help of SNN-SC, the transmission overhead when deploying complex AI models on vehicles through CI is significantly reduced.
\item We present a novel spiking neuron model called Integrate and Hybrid Fire (IHF) in SNN-SC, which is capable of outputting both spike and membrane potential information simultaneously. By leveraging IHF, the cloud can effectively recover the received semantic information into informative floating-point features, leading to improved task performance in the cloud.
\item We further enhance the performance of SNN-SC transmission features by maximizing the entropy of semantic information. By doing so, the mutual information between the transmitted and received semantic information is maximized, enabling the model to maintain high performance of cloud tasks under poor channel conditions.
\item Image classification and segmentation models are adopted to prove the effectiveness of SNN-SC. Experimental results reveal that the SNN-SC exhibits superior robustness and lower computational complexity than the existing SC models. In addition, SNN-SC has higher compression capability than the separated source and channel coding method, and overcomes the "cliff effect" of the separation method wherein performance drops sharply under poor channel conditions.
\end{enumerate}

The rest of this paper is organized as follows. Section \ref{sec:2} reviews some related works. Section \ref{sec:3} describes the architecture, spiking neurons and training method of the proposed SNN-SC. Experiments and ablation analysis are presented in Section \ref{sec:4}. Section \ref{sec:5} finally concludes this paper.

\section{Related works}
\label{sec:2}
\subsection{Collaborative Intelligence}
With the development of autonomous driving \cite{86}, sensors on vehicles generate billions of bytes of data every day \cite{81}. To fully analyze the data, it is imperative to apply AI models to vehicles. However, given the limited hardware computing resources on the vehicle, it is challenging to run complex AI models independently. To address this issue, CI has been proposed, which enables the execution of complex models through edge-cloud collaboration \cite{2}. In CI, the AI model is split into two parts, and intermediate features are transmitted from the edge device (such as a vehicle) to the cloud.

Selecting appropriate split points to balance the computing overhead of edge devices and cloud is a critical issue in CI. \cite{2} predicted each layer's latency/energy cost based on its type and configuration, and selected the optimal partition point of a given DNN. In \cite{80}, a reconfigurable DNN architecture is proposed to configure the optimal split point according to the edge-cloud CI environment. \cite{83} proposed an algorithm that can perform different model splitting schemes based on the computing resources of each user in a multi-user scenario. 

To reduce the transmission overhead of the intermediate features, many feature compression methods are proposed. In \cite{4} and \cite{5}, the feature was treated as a video with multiple frames, and compressed using HEVC-Intra and HEVC-Inter \cite{6} of the video coding framework. In \cite{7} and \cite{8}, the feature was rearranged to a tiled image, and compressed by image codecs PNG \cite{9} and JPEG \cite{10}. Apart from traditional approaches, the butterfly unit proposed in \cite{12} utilized Auto-Encoder (AE) \cite{11} to reduce the feature size. To further enhance model compression capability, the bottlenet unit \cite{13} used JPEG to compress the compression result of the AE. When the split point involves multiple features with different scales, the framework in \cite{14} first merged them into a large feature and then compressed it with AE. To get high task performance under low latency, an end-to-end approach is proposed in \cite{94}. It can achieve optimal resource management by jointly considering the edge devices' sensing, computation, and communication.

Split point selection and feature compression methods help reduce the cost of computation and  transmission edge devices. However, in real-world scenarios, wireless channels connecting edge devices to the cloud are imperfect. To mitigate the impact of channel noise, the semantic communication model can be used to transmit features.

\subsection{DNN-Based Semantic Communication}
\label{sec:2-2}
Nowadays, DNN-based semantic communications have shown great potential in extracting semantic information from various sources and achieving robust performance in bad channel environments. Existing works on DNN-based SC are mainly applied in three scenarios: image transmission, text transmission, and feature transmission. 

In image transmission, SC aims to reconstruct the image pixels. To this end, a multi-layer AE framework was proposed in \cite{17}, which modeled the AWGN and Rayleigh slow fading channels as non-trainable layers, thereby introducing random noise to the transmission process. By jointly optimizing the encoder and decoder under the noisy channel, the image can be robustly transmitted over the channel. The peak signal-to-noise ratio (PSNR) is a widely used measurement in image transmission, which describes the difference between the pixels of the reconstructed and original images. To improve the adaptability of the image SC model, \cite{18} proposed a new training strategy that trains the model under channels with randomly varying SNRs, and \cite{19} proposed a attention module that leverages SNR information to enhance the robustness of semantic encoding. Both of these studies enable the image SC model to adapt to channels with different SNRs. In a distributed scenario, \cite{20} designed an SC framework with two encoders, which transmits a pair of correlated images over independent channels and jointly recovers them by one decoder, leveraging the common information across two images.

In text transmission, the primary objective of SC is to recover the meaning of sentences. To this end, a DeepSC model was proposed by \cite{54}, which comprises a Transformer-based semantic encoder, an AWGN channel, and a Transformer-based decoder. To evaluate the performance of DeepSC at the semantic level, a novel measure known as sentence similarity based on BERT \cite{76} was designed. To make the DeepSC model more affordable for IoT devices, \cite{55} proposed a distributed version of DeepSC called L-DeepSC, which reduces model complexity by pruning redundant parameters and reducing weight resolution. Additionally, to mitigate the semantic errors caused by channel interference in sentences with varying lengths, \cite{56} and \cite{70} introduced a hybrid automatic repeat request (HARQ) mechanism.

In feature transmission, SC focuses on task performance, which is affected by the reconstructed feature at the receiver. \cite{21} and \cite{22} designed CNN-based feature SC models to transmit semantic information of the intermediate features of the image retrieval model and the classification model, respectively. Downstream task performance, such as classification accuracy, is typically used as a measure of feature transmission quality. \cite{23} and \cite{24} designed SC models for transmitting the features of the multi-task model and the multi-modal model through the AWGN channel. In addition, the information bottleneck (IB) framework has been leveraged to formalize the trade-off between the informativeness of encoded semantic information and inference performance \cite{66}. To reduce computational costs, filter pruning strategies was used on classification model \cite{65}, and branch networks were introduced for early exiting in CI scenarios \cite{67}. \cite{68} and \cite{72} studied collaborative semantic communication, where the features of correlated source data among different users are transmitted via a shared channel, then fused in the cloud to perform retrieval or identification tasks. 

Most SC models only consider analog channels as the output of DNN models is typically represented in floating-point values. To enable the application of DNN-based SC models to digital channels, additional quantization or sampling steps are required to convert the model outputs to binary bits \cite{22,92,93}. In this study, we investigate the feasibility of using the feature SC framework for CI under digital channels, which eliminates the need for explicit quantization.




\subsection{Spiking Neural Networks}
SNNs are originally inspired by the communication mechanism in the brain where neurons interact through discrete spikes \cite{59}. Spiking neurons serve as the fundamental computational unit of SNNs, receiving continuous-valued input and converting it into spikes. Typically, spiking neurons operate in three distinct steps: charge, fire, and reset \cite{27}. These three steps are executed sequentially at each time step. To illustrate the working process of a spiking neuron, we use the Integrate-Fire (IF) neuron as an example \cite{36}.

During charging, the IF accumulates the input of the previous layer and the membrane potential of the previous time step as the membrane potential of the current time step:
\begin{equation}
m_{i}^{t}=  m_{i}^{t-1} + I_{i-1}^{t}, \label{equ:1}
\end{equation}
where $m_{i}^{t}$ represents the membrane potential of $i^{th}$ neuron, and $I_{i-1}^{t}$ is the weighted input from previous layer at time step $t$. After charging, if the membrane potential exceeds the threshold, the neuron will output a spike at the current time step, which is called the fire step and define by
\begin{equation}
S_{i}^{t}= 
\begin{cases}
1,\quad m_{i}^{t}>V_{th},\\
0,\quad \text{otherwise}, \label{equ:2}
\end{cases}
\end{equation}
where $S_{i}^{t}$ of Eq. (\ref{equ:2}) is the output spike of neuron $i$ at time step $t$, and $V_{th}$ is the firing threshold of the neuron. After firing a spike, the membrane potential will be reset, and there are two reset methods: hard reset and soft reset. In hard reset, the membrane potential is directly set to the preset value $V_{reset}$ \cite{38}. On the other hand, the membrane potential is subtracted by the threshold $V_{th}$ in soft reset \cite{39}. What's more, if the membrane potential is less than the threshold, a `0' occurs and the membrane potential does not change during the reset step. This process can be described as
\begin{equation}
m_{i}^{t} = 
\begin{cases}
(1-S_{i}^{t})m_{i}^{t}+S_{i}^{t}V_{reset}\quad &\text{(hard reset)}. \\
m_{i}^{t}-S_{i}^{t}V_{th} \quad &\text{(soft reset)}. \label{equ:3}
\end{cases}
\end{equation}

In addition to the IF that outputs binary spikes, there are spiking neurons like membrane potential (MP) neuron \cite{40} that outputs membrane potential values after the charge step.

As the fire step is non-differentiable, various training methods have been proposed. The two main methods are ANN-to-SNN conversion \cite{ 43, 44, 30, 32} and backpropagation with surrogate gradient \cite{ 26, 40, 33, 75}. The ANN-to-SNN method involves training a neural network with ReLU function and then replacing it with spiking neurons. However, the performance of SNNs resulting from direct conversion is generally poor, and additional operations such as normalization \cite{45, 46} or threshold adjustment \cite{ 29, 31} are required. The backpropagation method involves a continuous relaxation of the firing, enabling backpropagation with a surrogate gradient. This approach allows for end-to-end training of SNNs and can achieve competitive performance with ANNs.

In recent years, there have been several studies exploring the use of SNN in communication. Specifically, \cite{87} and \cite{88} utilized SNN to compress event data generated by neuromorphic sensors and transmit the compressed data over analog channels for identification. Moreover, \cite{61} employed SNN to transmit features of an image-text retrieval model on BSC, but this study does not consider feature compression.



\begin{figure*}[!t]
\centering
\includegraphics[width=1.0\linewidth]{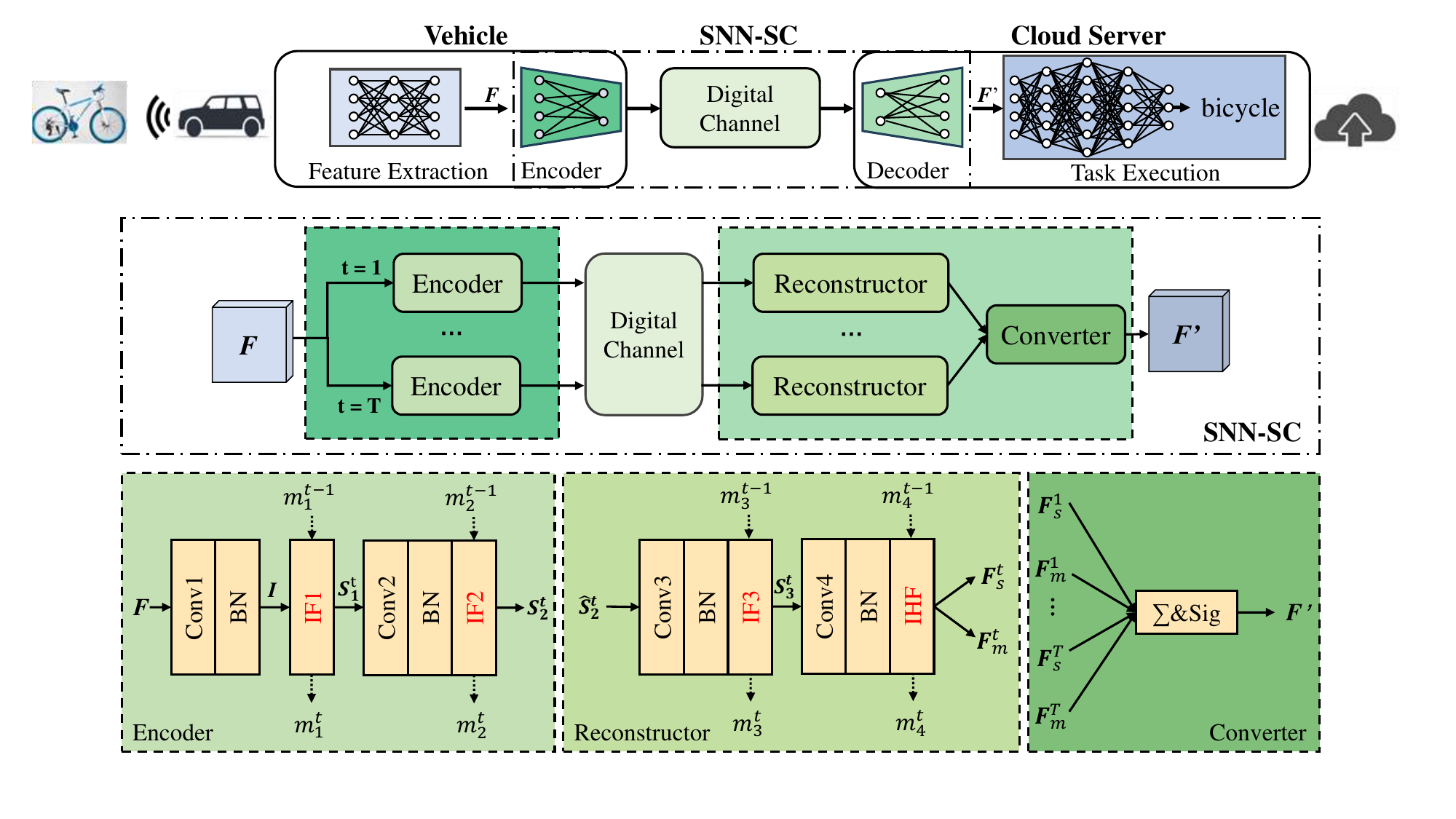}
\caption{The architecture of the SNN-SC model over digital channel.On the vehicle, the feature extraction module extracts original features from the input images. The encoder of SNN-SC extracts compact and robust semantic information from the original features, effectively reducing the required bandwidth of feature transmission. On the cloud server side, the reconstructor of SNN-SC decodes and recovers the received semantic information, and then the converter integrates the reconstructed results over multiple time steps. The final reconstructed features are used to perform downstream task.}
\label{fig:1}
\end{figure*}

\section{Methodology}
\label{sec:3}
\subsection{Overview of the whole model}
In Fig. \ref{fig:1}, an image classification model is applied to the autonomous driving scenario, where the model is split and deployed on the vehicle and cloud server. The feature extraction module on the vehicle analyzes the image captured by the sensor and extracts original features. Then SNN-SC encoder extracts semantic information from the feature and sends it to the cloud. The extracted semantic information consists of binary values and contains information related to the classification task. It can be directly applied to the digital channels without additional quantization. Moreover, the size of the semantic information is much smaller than that of the original features, which considerably saves the bandwidth.  Finally, the cloud first reconstructs the feature by SNN-SC decoder and then uses the task execution module to obtain the classification result. Specially, both the encoder and decoder of SNN-SC are lightweight, making the computation introduced by the SNN-SC negligible compared to the whole network.

The digital channels considered in this paper are BSC and BEC. In the BSC, random bit flips occur with a specific probability, known as the bit error rate. In the BEC, random bit erasures occur with a specific probability, known as the bit erase rate. The erased bits are then assigned a value of either 0 or 1 at equal probability during reception.

\subsection{Details of the SNN-SC}
\subsubsection{Encoder}
On the vehicle, the encoder filters out task-irrelevant information in the original features, preserving the compact and robust semantic information. We denote the original feature generated from the feature extraction model on the vehicle as $\boldsymbol{F}\in \mathbb{R}^{c\times h\times w}$, and $\boldsymbol{F}$ is first compressed by Conv1-BN module. In Conv1-BN, the convolutional layer (Conv1) is utilized to discover useful information and reduce the dimension of  $\boldsymbol{F}$ , then the batch normalization (BN) layer standardizes the features by subtracting the batch mean and dividing by the batch standard deviation.  This process can be written as follows:
\begin{equation}
\boldsymbol{I} =\mathrm{E_1}(\boldsymbol{F} ;\theta_1 ),
\end{equation}
where $\boldsymbol{I} \in \mathbb{R}^{c1\times h1\times w1} $ represents the preliminary compression result, and $\mathrm{E_1}(\cdot;\theta_1 )$ represents the Conv1-BN module with learnable parameters $\theta_1$.  Then $\boldsymbol{I}$ is transformed into spikes by the IF1 neuron:
\begin{equation}
\boldsymbol{S_1^t} = \mathrm{IF_1}(\boldsymbol{I} ;m_1^{t-1}),
\end{equation}
where $\boldsymbol{S_1^t} \in \{0,1\}^{c_1\times h_1\times w_1} (t \in \{1,...,T\})$ represents the transformed result at the time step $t$, and $\mathrm{IF_1}(\cdot ;m_1^{t-1} )$ represents the IF1 neuron with membrane potential value of $m_1^{t-1}$. It is worth noting that the transform process is not only related to the input at the current time step, but also to the membrane potential state at the preceding time step, which utilizes the temporal dependencies to generate more robust results. Moreover, during the running of SNN-SC, $\boldsymbol{I}$ can be considered as a constant charge value and the membrane potential of IF1 is updated implicitly at each time step according to Eq. (\ref{equ:1}), Eq. (\ref{equ:2}), and Eq. (\ref{equ:3}), resulting in the corresponding update of $\boldsymbol{S_1^t}$. To get more compact and robust semantic information, $\boldsymbol{S_1^t}$ is further compressed by Conv2-BN module and transformed by IF2 neuron:
\begin{equation}
\boldsymbol{S_2^t} = \mathrm{IF_2}(\mathrm{E_2}(\boldsymbol{S_1^t} ;\theta_2 );m_2^{t-1}),
\end{equation}
where $\boldsymbol{S_2^t} \in \{0,1\}^{c_2\times h_2\times w_2}$ represents the extracted semantic information at time step $t$ with dimension $c_2 * h_2  * w_2 < c_1 * h_1 * w_1 <c * h * w$, $\mathrm{E_2}(\cdot;\theta_2 )$ represents the Conv2-BN module with learnable parameters $\theta_2$, and $\mathrm{IF_2}(\cdot ;m_2^{t-1})$ represents the IF2 neuron with membrane potential of $m_2^{t-1}$.

The extracted semantic information $\boldsymbol{S_2^t} $ is directly transmitted to the cloud through digital channels such as BSC or BEC. The transmission process is rewrite as follows:
\begin{equation}
\boldsymbol{\hat{S_2^t}} = \eta(\boldsymbol{S_2^t}; p), \label{equ:8}
\end{equation}
where $\eta(\cdot; p)$ represents the digital channel model with the error probability $p$ of each bit in the bitstream, and $\boldsymbol{\hat{S_2^t}}$ is the disturbed version of $\boldsymbol{S_2^t}$. Since $\boldsymbol{\hat{S_2^t}}$ consists of binary spikes, the amount of data transmitted on the channel at each time step is $c_2* h_2 * w_2 $ bits.

\subsubsection{Decoder} On the cloud, the decoder consists of a reconstructor and a converter, which fully recovers the semantic information to ensure efficient reasoning for subsequent classification task. At each time step, the reconstructor receives the disturbed semantic information and decompresses it into the results with higher dimensions. Similar to the encoder, there are two modules in the reconstructor. At time step $t$, $\boldsymbol{\hat{S_2^t}}$ is first decompressed by Conv3-BN module. In Conv3-BN, convolutional layer is utilized to filter noise in semantic information and increase the dimension of semantic information. Then IF3 further denoises and transforms the results using time dependency. This process can be written as follows:
\begin{equation}
\boldsymbol{S_3^t} = \mathrm{IF_3}(\mathrm{R_1}(\boldsymbol{\hat{S_2^t}};\phi_1);m_3^{t-1}),
\end{equation}
where $\boldsymbol{S_3^t} \in \{0,1\}^{c_1\times h_1\times w_1}$ represents the preliminary decompressed result, $\mathrm{R_1}(\cdot;\phi_1)$ represents the Conv3-BN module with learnable parameters $\phi_1$, and $\mathrm{IF_3}(\cdot ;m_3^{t-1})$ represents the IF3 neuron with membrane potential value of $m_3^{t-1}$. 

Then $\boldsymbol{S_3^t}$ is further decompressed Conv4-BN module and transformed by IHF neuron:
\begin{equation}
\boldsymbol{F_s^t}, \boldsymbol{F_m^t} = \mathrm{IHF}(\mathrm{R_2}(\boldsymbol{S_3^t};\phi_2);m_4^{t-1}), \label{equ:9}
\end{equation}
where $\boldsymbol{F_s^t} \in \{0,1\}^{c\times h\times w}$ and $\boldsymbol{F_m^t} \in \mathbb{R}^{c\times h\times w}$ represent the final decompression results of reconstructor, $R_2(\cdot;\phi_2)$ represents the Conv4-BN module with learnable parameters $\phi_2$, and $\mathrm{IHF}(\cdot ;m_4^{t-1})$ represents the IHF neuron with membrane potential value of $m_4^{t-1}$. It is worth mentioning that IHF is a novel spiking neuron proposed in this study, capable of outputting two types of information. Therefore, at each time step, the reconstructor obtains two different decompression results, one containing spike information and the other containing membrane potential information. A comprehensive explanation of IHF will be provided in the next section.

During the iterative running of SNN-SC, the model parameters in the encoder and reconstructor are shared across all time steps. Once the encoder and reconstructor have completed $T$ time steps, a total of 2$T$ decompression results are obtained. Since the decompression results obtained at different time steps contribute differently to the final reconstructed feature, the converter model is proposed to adaptively fuse all decompression results instead of simply adding them. In the converter, a fully connected network (FCN) is used to assign different weights to all decompression results, and the weights can be adaptively optimized through end-to-end training. The activation function used is Sigmoid. The integrate process of converter can be written as follows:
\begin{equation}
\boldsymbol{F}' =\mathrm{FCN}([\boldsymbol{F_s^1}, \boldsymbol{F_m^1}, ..., \boldsymbol{F_s^T}, \boldsymbol{F_m^T}];{\psi}), \label{equ:10}
\end{equation}
where $\boldsymbol{F'}\in \mathbb{R}^{c\times h\times w}$ represents the final reconstructed feature, which has the same dimension as original feature $\boldsymbol{F}$. $ \mathrm{FCN}([\cdot];{\psi})$ indicates the converter with parameters ${\psi}$.  $\boldsymbol{F'}$ will be fed into the task execution network to perform image classification task.

Overall, SNN-SC is the first innovative attempt to apply SNN to digital semantic communication. By ingeniously utilizing the semantic extraction capability of neural networks and the analog-to-digital conversion capability of spiking neurons, it proves the compatibility of SNN with digital channels, providing experience and a research foundation for subsequent research in the field of digital semantic communication.

\subsection{Integrate and Hybrid Fire neuron (IHF)}
Based on our observation, both the IF that output binary spikes and the MP \cite{40} that outputs membrane potential values can only output limited information,  which limits the representation capacity of neurons. Therefore, this paper proposes a novel spiking neuron by combining the working principles of these two neurons.

To make full use of the spike information and membrane potential information of spike neurons, we propose an IHF neuron that outputs both spike and membrane potential information. Unlike existing SNN neurons, IHF has four working steps. The first three working steps of the IHF neuron are the same as the IF neuron, and then IHF adds another output step after the reset step. The added step outputs the reset membrane potential information, so the representation ability of IHF neurons is greatly enhanced, effectively overcoming the limitations of IF and MP. The work steps of IHF is modeled as:
\begin{subequations}\label{equ:6}
\begin{align}
m^{t}_i &=  m^{t-1}_i + I^t_{i-1},  \label{equ:6a}\\
S^{t}_i &= 
\begin{cases}
1,\quad m^{t}>V_{th},\\
0,\quad \text{otherwise}, 
\end{cases}  \label{equ:6b}\\
m^{t}_i &= m^{t}_i-S^{t}_iV_{th}, \label{equ:6c} \\
M^t_i  &= m^t_i, \label{equ:6d}
\end{align}
\end{subequations}
where $m^t_i$ and $I^t_{i-1}$ are the membrane potential value and input from the previous layer at time step $t$, and $V_{th}$ is the firing threshold of the IHF neuron. In particular, $S^t_i$, and $M^t_i$ are the spike and membrane potential outputs of the IHF neuron at the time step $t$. Eq. (\ref{equ:6a}), Eq. (\ref{equ:6b}), Eq. (\ref{equ:6c}) and Eq. (\ref{equ:6d}) represent the charge, fire, reset, and output membrane potential steps of IHF, respectively. It is worth noting the step of IHF outputting membrane potential is not exactly the same as that of MP. Since MP only transmits information through the output membrane potential value and omits the fire step, its output is the membrane potential value after charging. In contrast, IHF decompose the charged membrane potential value into spike information and reset membrane potential information through the fire step, which makes more detailed use of the effective information hidden in the neuron.

\begin{figure}[h]
\centering
\includegraphics[width=0.85\linewidth]{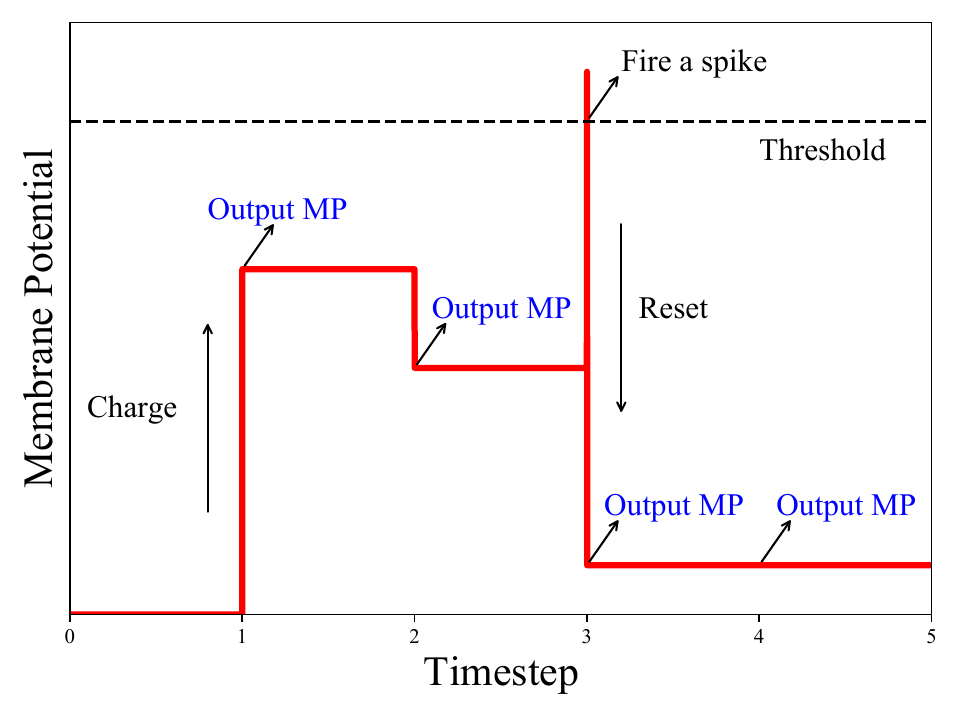}
\caption{The change of IHF membrane potential.}
\label{fig:2}
\end{figure}

The change of IHF membrane potential with soft reset over time steps is shown in Fig. \ref{fig:2}. Similar to the IF, if the membrane potential of the IHF exceeds the threshold, the IHF will output a spike, and the membrane potential will be reset. Notably, regardless of whether a spike is output, the IHF will output the membrane potential value (i.e., Output MP) at each time step, which fully compensates for the information loss caused by only outputting spike information. In the reconstructor, IHF is used to decode the received semantic information, and two outputs are generated, which provide rich information for the converter.

By flexibly exploiting the spike firing mechanism inherent in spiking neurons to dissect a single type of input information into distinct binary spike information and float membrane potential information, IHF makes the final reconstructed features contain more detailed task-related information. In addition, IHF also inspires the study of SNNs, prompting researchers to recognize and emphasize the significance of the information emission mechanisms within spiking neurons.

\subsection{Maximizing the entropy of semantic information}
In addition to the structural design of the SNN-SC model, we also improve the transmission efficiency of the model from the perspective of the mutual information of channel inputs and outputs, which has not been considered in existing digital semantic communication research. The mutual information between two random variables measures the correlation between them, and higher mutual information means a higher correlation. In this paper, we aim to maximize the mutual information between channel inputs and outputs,  ensuring that the semantic information received by the cloud is most relevant to the original semantic information.

Consider a BSC with probability $\epsilon$ of incorrect transmission, $X \in \{0,1\}^N$ and $Y \in \{0,1\}^N$ are the input and output of the channel respectively, where $p_0$ and $q_0$ are the probabilities of 0 in $X$ and $Y$. The problem is formulated as
\begin{equation}
\max \limits_{p_0} I(X;Y) = \max \limits_{p_0} H(Y) - H(Y|X),
\end{equation}
where
\begin{equation}
\begin{cases}
\hfil H(Y|X)= \epsilon\log_2{(1/\epsilon)}+(1-\epsilon)\log_2{(1/(1-\epsilon))},\\
\hfil H(Y)= q_0\log_2{(1/q_0)}+q_1\log_2{(1/q_1)},\\
\hfil q_0= p_0(1-\epsilon)+p_1\epsilon,\\
\hfil q_1= p_0\epsilon+p_1(1-\epsilon).\\
\end{cases}
\end{equation}

It shows that $H(Y|X)$ does not depend on the input distribution, and it is a constant value when $\epsilon$ is known. On the other hand, $H(Y)$ is maximized when $q_0 = q_1 = 1/2$ which happens when $p_0 = p_1 = 1/2$. Therefore, $I(X;Y)$ is maximized when 0 and 1 are transmitted with equal probability and reaches the channel capacity $1-H(\epsilon)$ \cite{89}. 

To make the 0 and 1 in the semantic information extracted by SNN-SC evenly distributed, we designed a loss based on maximizing information entropy for model training:
\begin{equation}
\mathcal{L}_\text{entropy} = (\mathbb{\alpha} - (p_0\log_2(1/p_0)+p_1\log_2(1/p_1)))^2, \label{equ:4}
\end{equation}
where $p_0$ and $p_1$ are estimated by the frequency of 0 and 1 in semantic information $\boldsymbol{S_2^t}$. SNN-SC can be trained end-to-end by using the surrogate gradient \cite{26}, and the entropy of semantic information will gradually approach the value of $\alpha$ during training. In this paper, $\alpha$ is set to 1.0 to maximize the information entropy of the semantic information, so that 0 and 1 in the semantic information are equally distributed.

By maximizing the semantic information entropy, the useful information contained in the semantic information received by the cloud reaches its maximum value. Consequently, the information transmission efficiency of the model is improved, and the task performance of the cloud is also improved.

\subsection{Training Strategy}
As ResNet50 (Residual Network) \cite{48} has shown impressive image classification abilities and is comprised of several bottleneck blocks linked in series, it is conducive for model segmentation and edge-cloud deployment. This makes it an ideal foundation for introducing the model training strategy. Since directly training ResNet50 combined with SNN-SC from scratch shows slow convergence \cite{22}, we propose a three-step training strategy to comprehensively train the entire model. Similar training strategies can also be used when SNN-SC is applied to other models.

The first step is to train ResNet50 using the cross-entropy (CE) loss function to reach the desired accuracy of the task:

\begin{equation}
\mathcal{L}_\text{CE} = -\frac{1}{N}\sum_{i=1}^{N} y_{i} \log(\hat{y}_{i}) ,
\end{equation}
where $N$ represents the total number of categories, $ y_{i}$ and $\hat{y}_{i}$ represent the label and the the prediction result of ResNet50 respectively. Once ResNet50 has converged, we select an intermediate feature as the split point, effectively dividing ResNet50 into two parts: the vehicle part and the cloud part. Then SNN-SC is established at the split point to extract and transmit semantic information through digital channels like Fig. \ref{fig:1}, and SNN-SC will run iteratively for $T$ time steps.

The second step is to train the SNN-SC framework while maintaining the parameters of ResNet50 fixed. The loss function is formulated by combining the cloud classification task loss and the entropy maximization loss in Eq. (\ref{equ:4}):
\begin{equation}
\mathcal{L}_\text{total}= \mathcal{L}_\text{CE} + \mathcal{L}_\text{entropy}. \label{equ:5}
\end{equation}

The third step is to jointly fine-tune the whole model. The whole model is trained end-to-end, and the loss function of this step is the same as the second step, but with a lower learning rate.

As mentioned earlier, SNN-SC is a task-oriented SC framework, so we focus on the performance of tasks in the cloud like classification accuracy, not whether the decoder reconstructs the original features perfectly. In this case, the loss function does not consider the symbol-level distance between the original features and the reconstructed features. 
\section{Experiments and Results}
\label{sec:4}
\subsection{Experiment Setup}
In this paper, ResNet50 \cite{48} and CCNet \cite{90} are deployed in the CI scenario, where ResNet50 performs image classification and CCNet performs semantic segmentation, both of which can be used in autonomous driving. ResNet50 is trained on CIFAR-100 \cite{49} and CCNet is trained on FoodSeg103 \cite{91}.  In CIFAR-100, the training set contains 50,000 images and the testing set contains 10,000 images, with 100 image category annotations. After training for 100 epochs with a learning rate of 1e-4, ResNet50 achieves a classification accuracy of 77.81\%. Additionally, in FoodSeg103, the training set contains 4,983 images and the testing set contains 2,135 images, with 103 semantic segmentation category annotations. After training for 150 epochs with a learning rate of 1e-4, CCNet achieves a mean Intersection over Union (mIoU) score of 34.1\%. During training, both models used the cross-entropy loss function and the Adam optimizer. In addition, ResNet50 and CCNet are built by Pytorch \cite{50}, and SNN-SC is built by SpikingJelly \cite{51}, an open-source deep learning framework for SNN. Training and inference processes are performed on Tesla V100 GPUs.

\subsection{Baseline Methods}
Two kinds of baseline methods used in this paper: the DNN-based SC frameworks, and the typical separate source and channel coding.
\subsubsection{DNN-based SC framework}
\begin{figure}[h]
\centering
\includegraphics[width=\linewidth]{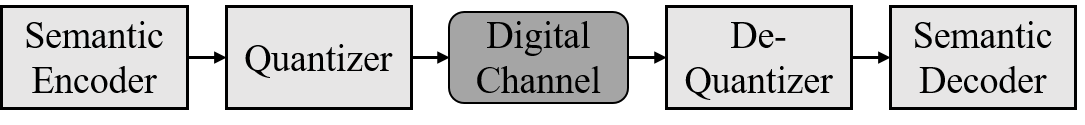}
\caption{The general architecture of the digital SC framework.}
\label{fig:3}
\end{figure}

A general framework of an existing SC model that runs on the digital channel is shown in Fig. \ref{fig:3}, and the model is mainly divided into a semantic encoder, a quantizer, a dequantizer, and a semantic decoder. The semantic encoder extracts the semantic information and the quantizer converts the semantic information into a bit sequence. After passing through the digital channel, the dequantizer converts the received bit sequence into the floating-point semantic information, and the semantic decoder restores it into the reconstructed feature.

The semantic encoder and decoder are constructed by DNNs, while the quantizer and dequantizer are achieved in several ways in existing SC models. In \cite{22}, uniform quantization is utilized. A trained non-linear quantization is proposed in \cite{92}. Moreover, in \cite{93}, the bit sequence is obtained by performing Bernoulli sampling by using the extracted semantic information as a probability. These models are used as baseline methods in this paper.

\subsubsection{Separate source and channel coding}
To perform the source and channel coding separately, we use the Joint Photographic Experts Group (JPEG) for source coding and the convolutional code \cite{78} for channel coding. Features are first quantized into 8-bit integers, then compressed by JPEG, and finally channel-coded by convolutional codes. The encoding results are transmitted over the digital channel, and the receiver performs corresponding channel decoding, source decoding, and dequantization operations.

\begin{table}
\normalsize
\caption{The parameter setting of the SNN-SC}
\label{tab:1}
\centering
\renewcommand\arraystretch{1.3}{
\begin{tabular}{c|c|c}
\hline
\hline
    \textbf{Module}    & \textbf{Layer} & \textbf{Parameters}\\
\hline
\hline
\multirow{2}*{\textbf{Encoder}}& Conv1 & {\makecell[c]{ kernel size = 3, output = $C$/8\\stride = 1}}\\
                     \cline{2-3}
                    ~& Conv2  &\makecell[c]{ kernel size = 3, output = $C$/64\\stride = 1}\\
\hline
\multirow{2}*{\textbf{Reconstructor}}& Conv3 & \makecell[c]{ kernel size = 3, output = $C$/8\\stride = 1}\\
                     \cline{2-3}
                    ~& Conv4  & \makecell[c]{ kernel size = 3, output = $C$\\stride = 1}\\
\hline
          \textbf{Converter} & FCN   & input = 2$T$, output = 1\\
\hline

\end{tabular}
}
\end{table}

\subsection{Implementation Details}
\label{sec:4-3}
When deploying the ResNet50 and CCNet under CI, the split points are the output of the 14th bottleneck block in ResNet50, which has feature dimensions of (2048, 4, 4), and the output of the backbone model in CCNet, which has feature dimensions of (2048, 65, 65). The parameter settings for the encoder, reconstructor, and converter in the SNN-SC model are presented in Table \ref{tab:1} for features with $C$ channels. Similarly, the baseline methods employ identical settings in their encoders and decoders.

According to the training strategy, after the trained ResNet50 and CCNet are split between vehicle and cloud, SNN-SC is inserted at the split point and trained. When transmitting ResNet50 features, SNN-SC is first trained for 50 epochs with a learning rate of 1e-4, and then the entire model is fine-tuned for 50 epochs with a learning rate of 1e-5. When transmitting CCNet features, SNN-SC is first trained for 60 epochs with a learning rate of 1e-4, and then the entire model is fine-tuned for 60 epochs with a learning rate of 1e-5. The loss function is Eq. (\ref{equ:5}), and $\alpha$ is set to 1.0. The firing thresholds of the IF and IHF are both set to 1.0, and the reset methods are soft reset.  To enable SNN-SC to be trained end-to-end, the Sigmoid function $\sigma(x) =(1 + e^{-x})^{-1}$ is used as the surrogate function for the firing step in backpropagation. The training details of the baselines are the same, except for the loss function. Since the binarization process of quantization is non-differentiable, we cannot constrain the information entropy of the binarized bitstream in baselines. Therefore, we use the end-to-end task loss to train baselines in all training steps.

Furthermore, during training, the noisy digital channel is placed between the encoder and decoder of SNN-SC as a non-trainable layer, to randomly flip or erase the semantic information. In backpropagation, the interference of digital channel is treated as an identity mapping with a gradient of 1.

\subsection{Model Performance}

We use SNN-SC and baseline methods to transmit features of ResNet50 and CCNet, and classification accuracy and mIoU are used as metrics to measure the transmission performance of the models. As mentioned earlier, this paper uses the methods in \cite{22},\cite{92}, \cite{93}, and the separate coding scheme as baselines. To ensure fairness, the output feature dimensions and required bandwidths of the encoders in all SC models are the same. For example, if \cite{22} uses $n$-bit quantization, then the SNN-SC will run $n$ (i.e., $T$ = $n$) time steps to transmit the same number of bits. To verify the performance of the proposed SNN-SC under different scenarios, we trained multiple sets of models using various bandwidths ($T$ = 8, 4) and different channels (BSC, BEC). Subsequently, we evaluated their performance under diverse channel conditions, as illustrated in Fig. \ref{fig:4} and Fig. \ref{fig:9}. The separate source and channel coding is directly applied to split point of the model. Furthermore, during performance evaluation, the semantic communication process were repeated 10 times to mitigate the effect of randomness introduced by the digital channel.

In the figures, SNN-SC, CNN-Uni, CNN-NonUni, CNN-Bern, and JPEG+Conv represent the performance of the classification/segmentation model combined with SNN-SC, \cite{22},\cite{92}, \cite{93} and the separate coding, respectively.

\subsubsection{Model Performance on ResNet50}

\begin{figure*}[!t]
\centering
\includegraphics[width=0.75\linewidth]{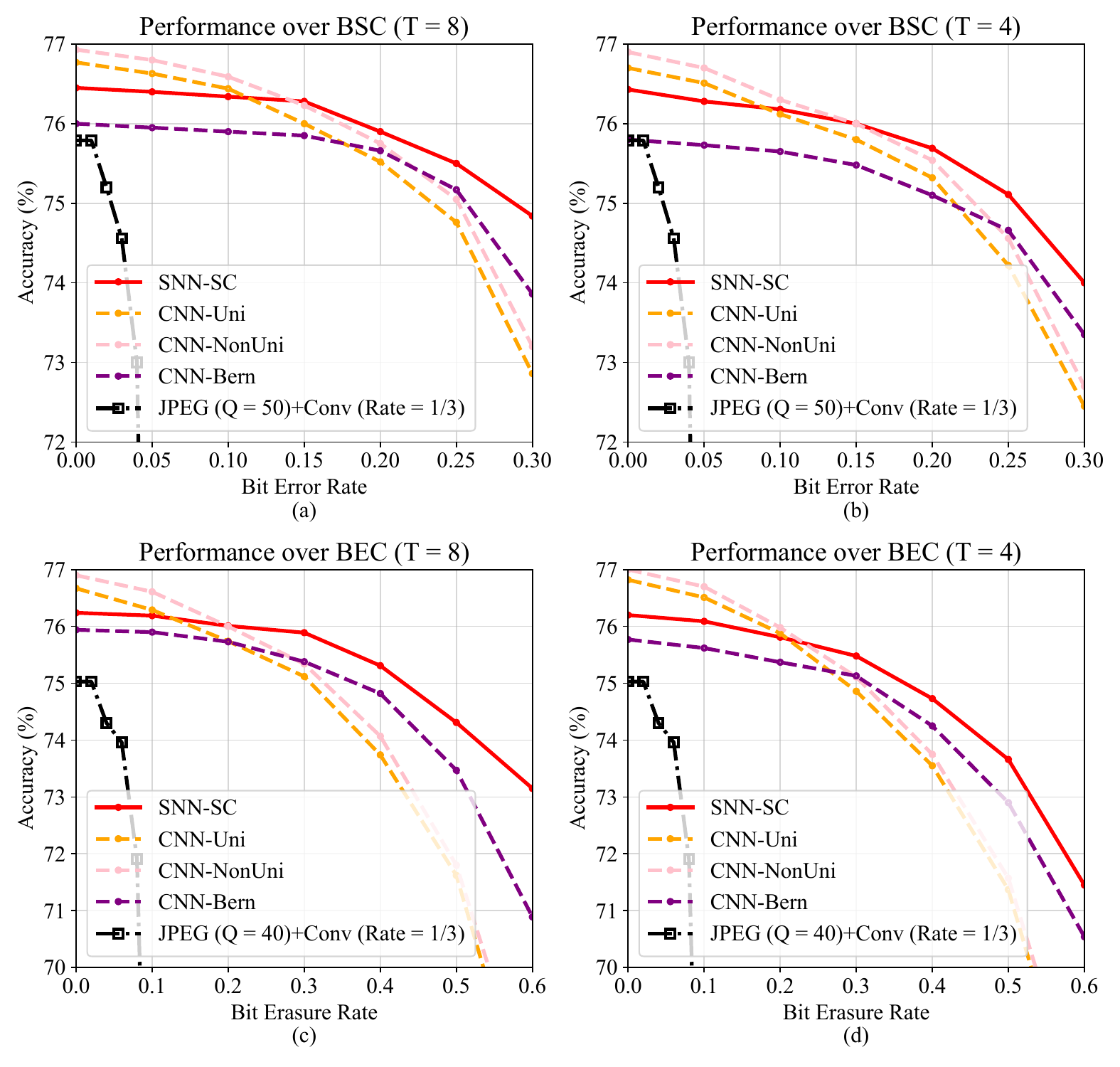}
\caption{Performance of SNN-SC and baselines when transmitting ResNet50 features over digital channels.}
\label{fig:4}
\end{figure*}

We study the performance of SC models on ResNet50 in Fig. \ref{fig:4}. On BSC, two groups of models are trained under different bandwidths (i.e., ${T}$ = 4 and 8), and the bit error rates of training are randomly sampled from [0, 0.3]. Then we test the accuracy of these models on a range of bit error rates in Fig. \ref{fig:4}(a) and Fig. \ref{fig:4}(b). 

\label{sec:4-4}

The results indicate that when the test bit error rate is 0, meaning that there is no noise interference in the transmission process, the classification accuracy of all methods is slightly lower than the original ResNet50. This is due to information loss during feature compression and decompression. As the bit error rate increases, the performance of SNN-SC and baselines degrades due to noise interference. When the bit error rate is lower than 0.1, indicating good channel conditions, the performance of SNN-SC is very close to that of the baseline, and the performance gap between SNN-SC and the baseline is less than 0.5\%, which is acceptable. However, when the channel bit error rate is larger than 0.1, indicating poor channel conditions, the performance of CNN-Uni and CNN-NonUni drops significantly, while that of SNN-SC and CNN-Bern decreases slowly. When the bit error rate is larger than 0.15, SNN-SC outperforms all the baselines.

The performance of CNN-Uni and CNN-NonUni drops significantly because the bits in both uniform and non-uniform quantization results have different weights. Bit errors with high weights introduce large errors, leading to significant performance losses, especially under poor channel conditions. Although the performance of CNN-Bern decreases slowly, the Bernoulli sampling-based quantization method loses too much information, resulting in the model's performance being consistently lower than that of SNN-SC. In contrast, SNN-SC uses spiking neurons to convert floating-point numbers into spikes, and the spikes are treated equally. Moreover, the spiking neurons only fire when the membrane potential exceeds the threshold, which also helps the decoder filter out some membrane potential errors caused by channel interference.

To achieve comparable performance to SNN-SC, the JPEG with a quality value of 50 (i.e., $Q$ = 50) and convolutional code with a code rate of 1/3 (i.e., $Rate$ = 1/3) is used in JPEG+Conv. The results show that JPEG+Conv suffers from the `cliff effect', that is, the performance drops sharply when the bit error rate is greater than 0.05. The reason for the `cliff effect' is that the channel decoding cannot correct errors normally in worse channel conditions, resulting in a decrease in the quality of the reconstructed features obtained by source decoding, thus affecting the performance of downstream tasks. While SNN-SC overcomes the cliff effect and outperforms JPEG+Conv, especially under poor channel conditions. The behavior of SNN-SC and baselines are similar in $T$ = 8 and $T$ = 4 scenarios. Specifically, the reduction of transmitted data leads to a decrease in the models' robustness, resulting in lower classification accuracy under $T$ = 4 compared to $T$ = 8, particularly in scenarios with poor channel conditions.

On the other hand, two groups of models are trained under different bandwidths (i.e., ${T}$ = 8 and 4) on BEC, and the bit erasure rates of training are randomly sampled from [0, 0.3]. The test performance is shown in Fig. \ref{fig:4}(c) and Fig. \ref{fig:4}(d). Since the channel properties of BEC are different from those of BSC, the erased bit is randomly set to 0 or 1 before decoding at the receiver. Therefore, only half of the bits that are erased will be flipped. In this case, when the erasure rate of BEC is the same as the bit error rate of BSC, BEC has less impact on the bit sequence than BSC. In the figures, all the models are tested on a wider range of bit erasure rates (i.e., 0-0.6), and the performance degrades significantly only when the bit erasure rate is greater than 0.3. Overall, similar results can be observed, the performance of SNN-SC is close to CNN-Uni and CNN-NonUni at low bit erasure rates, and much better at high bit erasure rates. Moreover, SNN-SC consistently outperforms CNN-Bern and JPEG+Conv.

\begin{table}[h]
\caption{Required bandwidth and Compression Ratio of \\JPEG+Conv and SNN-SC on ResNet50.}
\centering
\resizebox{\linewidth}{9.5mm}{
\renewcommand\arraystretch{1.15}{
\begin{tabular}{c|c|c|c|c}
\hline
\hline
\multirow{2}{*}{\diagbox{Metric}{Method}}&\multicolumn{2}{c|}{JPEG+Conv} &\multicolumn{2}{c}{SNN-SC}\\
\cline{2-5}
      & Q = 50 & Q = 40 & T = 8 & T = 4 \\
\hline
\hline
Bandwidth (bits)  &1.3$\times$10$^5$ & 1.1$\times$10$^5$ & 4.1$\times$10$^3$ & 2.0$\times$10$^3$ \\
\hline
Compression Ratio & 8.2$\times$   & 9.8$\times$   & \textbf{256$\times$}  & \textbf{512$\times$}   \\
\hline
\end{tabular}
}}
\label{tab:4}
\end{table}

To demonstrate the compression capability of SNN-SC, Table \ref{tab:4} shows the required bandwidth and the compression ratios of SNN-SC and JPEG+Conv, when transmitting ResNet50 features. The dimension of the original feature is (2048, 4, 4), and each element is a 32-bit floating point number. The compression ratio compares the number of bits contained in the original feature and the number of transmitted bits. It can be seen that when T = 8 and 4, the compression ratios of SNN-SC for features are 256$\times$ and 512$\times$, respectively, which are much higher than the compression ratios of JPEG+Conv. Therefore, compared with directly transmitting the original features or using the JPEG+Conv scheme, the proposed SNN-SC dramatically reduces the required bandwidth resources during transmission.

\subsubsection{Model Performance on CCNet}
\begin{figure*}[!t]
\centering
\includegraphics[width=0.75\linewidth]{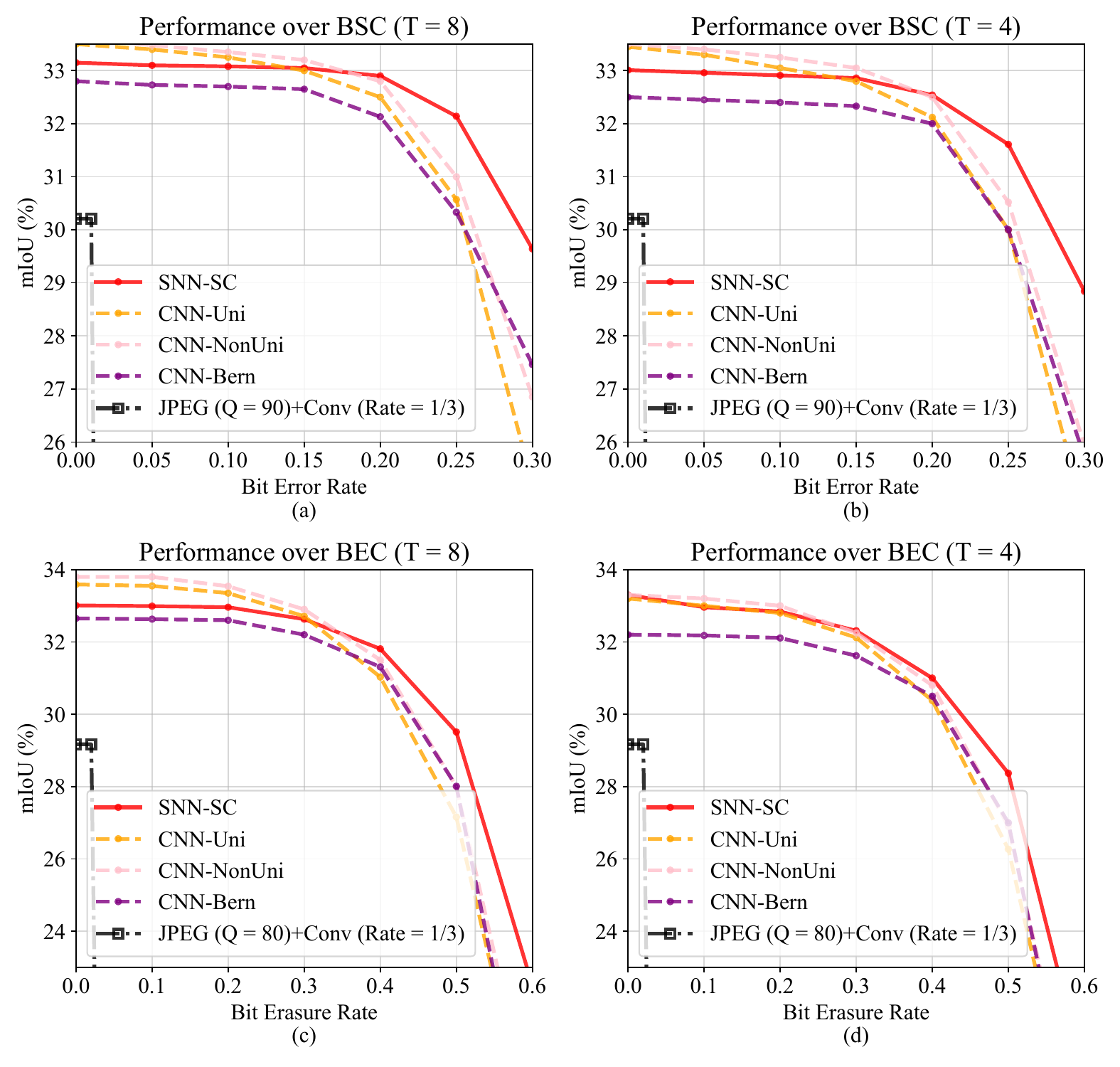}
\caption{Performance of SNN-SC and baselines when transmitting CCNet features over digital channels.}
\label{fig:9}
\end{figure*}

We study the performance of SNN-SC on CCNet in Fig. \ref{fig:9}. SNN-SC and baselines are trained under the same conditions as before. As shown in Fig. \ref{fig:9}, when the channel conditions are good, the mIoU of SNN-SC is very close to the baselines. When the channel conditions are poor, the performance of SNN-SC is much better than the baselines on both BSC and BEC. Moreover, the performance of JPEG+Conv is much worse than SNN-SC, and suffers from the `cliff effect'.  When $T$ decreases, the robustness of all models decreases due to the reduction of transmitted bits. Overall, the behavior of SNN-SC and baselines in Fig. \ref{fig:9} is similar to that in Fig. \ref{fig:4}. In addition, Table \ref{tab:5} shows that SNN-SC achieves much higher compression ratios than JPEG+Conv, when transmitting CCNet features with dimensions (2048, 65, 65). 

\begin{table}[h]
\caption{Required bandwidth and Compression Ratio of \\JPEG+Conv and SNN-SC on CCNet.}
\centering
\resizebox{\linewidth}{9.5mm}{
\renewcommand\arraystretch{1.15}{
\begin{tabular}{c|c|c|c|c}
\hline
\hline
\multirow{2}{*}{\diagbox{Metric}{Method}}&\multicolumn{2}{c|}{JPEG+Conv} &\multicolumn{2}{c}{SNN-SC}\\
\cline{2-5}
      & Q = 90 & Q = 80 & T = 8 & T = 4 \\
\hline
\hline
Bandwidth (bits)  & 2.6$\times$10$^7$ & 1.6$\times$10$^7$ & 1.1$\times$10$^6$ & 5.4$\times$10$^5$ \\
\hline
Compression Ratio & 10.7$\times$   & 17.3$\times$   & \textbf{256$\times$}  & \textbf{512$\times$}   \\
\hline
\end{tabular}
}}
\label{tab:5}
\end{table}

In summary, the experiments conducted demonstrate that the SNN-SC model proposed in this study is capable of generalizing across various bandwidths, channels, and models. Based on these findings, we can conclude that the SNN-SC framework is more suitable for feature SC on digital channels than SOTA SC models and the separate source and channel coding method. This is due to its superior robustness in poor channel conditions, as well as its higher feature compression ratio compared to separate coding. Moreover, the proposed SNN-SC is a plug-and-play feature transmission model, that can be applied to other models under CI without altering.

\subsection{Ablation Studies}
In this section, we analyze the impact of the proposed IHF neuron and maximizing entropy-based loss function on model performance. To ensure the persuasiveness of our experimental results, we adopt the method of controlling variables when conducting ablation experiments. Specifically, the model we use is ResNet50 and the digital channel is BSC. Furthermore, all other parts of the model, such as the training strategy and parameter settings of the spiking neurons, are kept constant except for the modules that we are analyzing.

First, experiments are conducted to investigate whether both outputs of IHF neurons contribute to the transmission efficiency of semantic information in SNN-SC. From Eq. (\ref{equ:6}) we can infer that, if the membrane potential output is removed, the IHF will degenerate into an IF neuron. On the other hand, we refer to an IHF neuron whose spike output is removed and only outputs membrane potential as an IHF-m neuron. Based on that, the SNN-SC framework with IHF, IF, and IHF-m neurons at the end of the reconstructor are referred as the IHF model, IF model, and IHF-m model, respectively. These models are trained using the CE function to eliminate the effect of maximizing entropy-based loss on the models, and the performance of these three models are evaluated by performing feature SC on ResNet50.
\begin{figure}[h]
\centering
\includegraphics[width=0.75\linewidth]{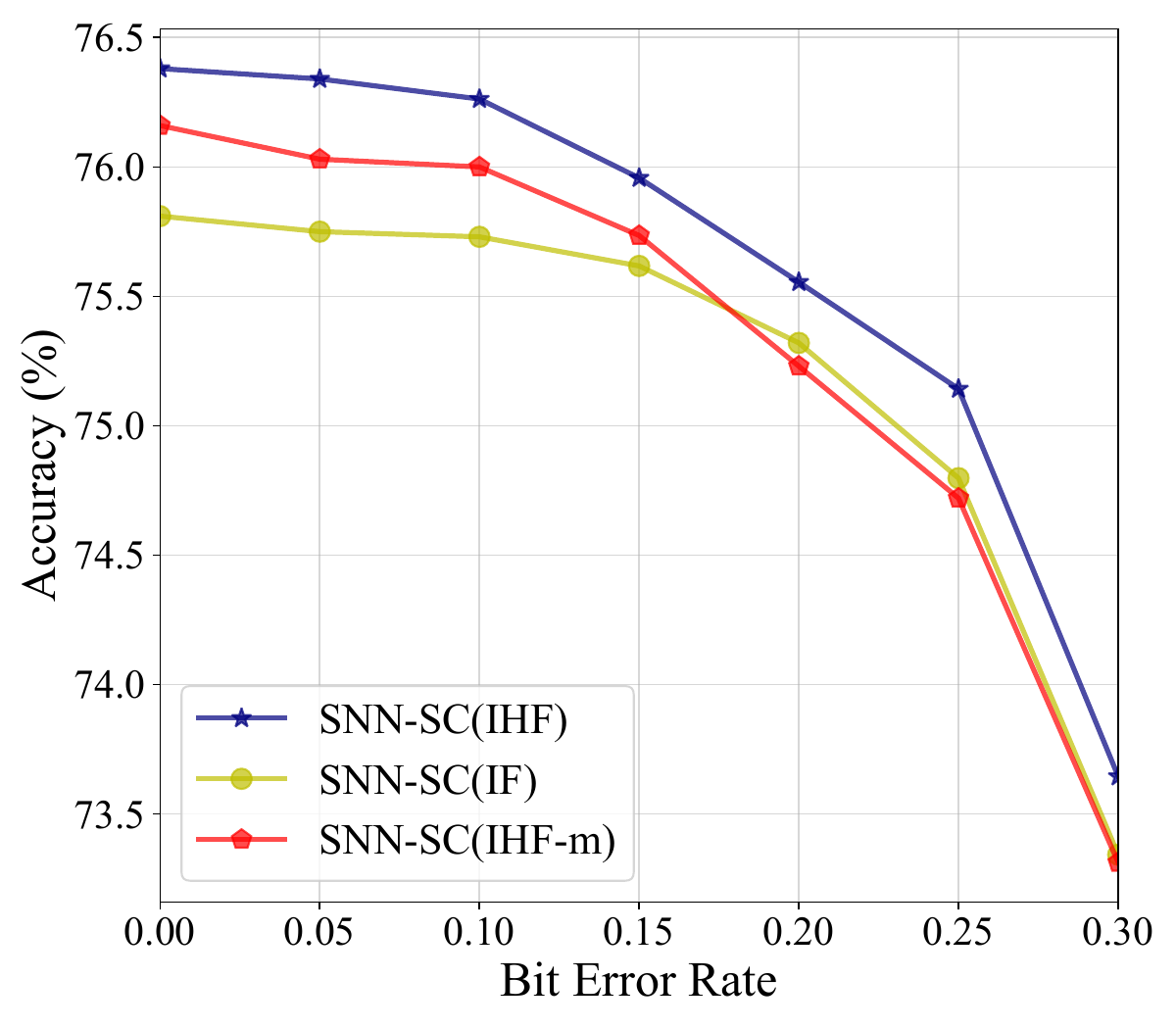}
\caption{Performance of IHF model, IF model, and IHF-m model.}
\label{fig:5}
\end{figure}

The test performance is illustrated in Fig. \ref{fig:5}. In the figure, SNN-SC(IHF), SNN-SC(IF), and SNN-SC(IHF-m) denote the test performance of ResNet50 with the IHF model, IF model, and IHF-m model, respectively. It is obvious that, in the test bit error rate range, the performance of SNN-SC(IHF) is much better than both SNN-SC(IF) and SNN-SC(IHF-m). The reason is that the output of IHF contains more information than IF and IHF-m. IHF fully exploits the internal membrane potential information and external spiking information of neurons, and enriches the input of the converter, thereby rendering the reconstructed features more conducive to subsequent classification task inference. Based on these experiments, we can conclude that IHF neurons have a more positive effect on feature SC compared to IF and IHF-m, and that both spike output and  membrane potential output contribute to the transmission performance of SNN-SC.

\begin{figure}[h]
\centering
\includegraphics[width=0.75\linewidth]{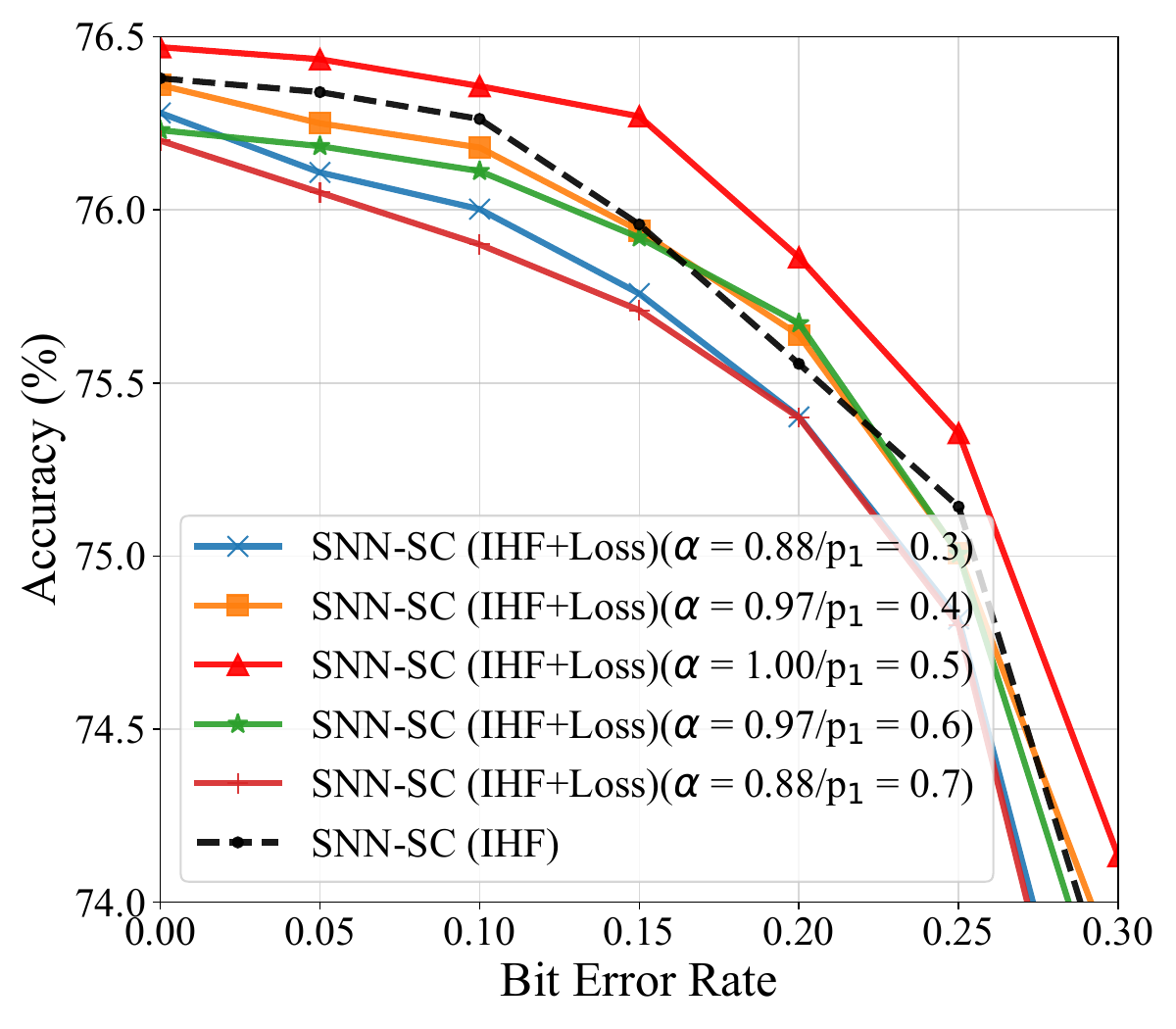}
\caption{Performance of the models trained with different $\alpha$ values in $\mathcal{L}_\text{entropy}$.}
\label{fig:7}
\end{figure}

Secondly, the effect of maximizing the entropy of the extracted semantic information (i.e., the output of the encoder) during model training are analyzed. Since IHF has been shown to have a positive effect on the model, the ablation experiments are based on the IHF model in the following part of this section. To explore the impact of entropy of the extracted semantic information on model performance, multiple IHF models are trained with different $\alpha$ values in the loss. According to Eq. (\ref{equ:4}), calculating the information entropy of semantic information requires the probability of 0 and 1 (i.e., $p_0$ and $p_1$), and $p_0+p_1=1$. We obtain the frequencies of 0 and 1 in the semantic information through statistics and then use them as approximate estimates of $p_0$ and $p_1$ respectively. The IHF model trained by Eq. (\ref{equ:5}) is called the IHF+Loss model. The test results of ResNet50 with different IHF+Loss models are presented in Fig. \ref{fig:7}.

In Fig. \ref{fig:7}, the dotted line is the test performance of the IHF model trained without $\mathcal{L}_\text{entropy}$, and the solid lines are the IHF+Loss models trained with different $\alpha$ values. The results show that the performance of the model improves as the value of $\alpha$ increases. The optimal performance and robustness of the model are achieved when $\alpha$ is set to 1.0. Correspondingly, the entropy of the semantic information is limited to a maximum value of 1.0, i.e., $p_0=p_1=1/2$. In this case, the mutual information between the channel input and output is maximized, so that the useful information received by the cloud is maximized, and the task performance is highest. Therefore, we can conclude that the proposed loss in Eq. (\ref{equ:4}) also contributes positively to the model.

\subsection{Complexity Analysis}
Besides model performance, model complexity is also a crucial metric for assessing model effectiveness. This section presents an analysis of the model parameters and computational complexity of both the proposed SNN-SC and three baseline models.

For model parameters, the number of parameters in convolution layer is 
\begin{equation} \label{equ:11}
\mathrm{Param_{Conv}} = K^2\times C_{in} \times C_{out},
\end{equation}
and for fully connected layer is
\begin{equation} \label{equ:12}
\mathrm{Param_{FCN}} = f_{in} \times f_{out},
\end{equation}
where $K$ is the kernel size, $C_{in}$($C_{out}$) is the number of the input(output) channels of the convolutional layer, and $f_{in}$($f_{out}$) is the number of input(output) neurons of the fully connected layer. 

For computational complexity, we need to count the operations in SNN-SC and baselines. For DNN, a multiplication and addition (MAC) compution takes place per operation. The total operation number of convolutional layer is 

\begin{equation} \label{equ:13}
\mathrm{OP_{Conv}} = 2 K^2 \times C_{in} \times C_{out} \times H_{out} \times W_{out},
\end{equation}
and for fully connected layer is
\begin{equation} \label{equ:14}
\mathrm{OP_{FCN}} = 2 f_{in} \times f_{out},
\end{equation}
where the factor 2 comes from the fact MAC contains two operations, $H_{out}$ and $W_{out}$ are the hight and width of the output feature. For SNN, only accumulated compution (AC) is required when an incoming spike is received \cite{44}. Moreover, the total number of AC operations in SNN also depends on the average fire rate of the input feature (i.e., the ratio of 1 in the input feature) and the running time step $T$ \cite{79}. The operation number of SNN-SC can be calculated by

\begin{equation} 
\begin{split}
\mathrm{OP_{SNN-SC}} =& \mathrm{OP_{Conv1}}+ (fr_1\times \mathrm{OP_{Conv2}}+fr_2 \times \mathrm{OP_{Conv3}} \\ &+ fr_3 \times \mathrm{OP_{Conv4}})\times \frac{1}{2} \times T + \mathrm{OP_{FCN}},\\
\end{split}
\end{equation}
where $fr_1$, $fr_2$ and $fr_3$ are the fire rate of the input features to the Conv2, Conv3, and Conv4 layers, and the factor $\frac{1}{2}$ comes from the fact that the operation number of AC is half of MAC. On the other hand, the inputs to the first layer (Conv1) and the converter (FCN) are floating-point features, so the operation number of these two parts are calculated based on Eq. (\ref{equ:13}) and Eq. (\ref{equ:14}).  In the experiments, the ($C$, $H$, $W$) is set to (2048, 4, 4) for transmitting ResNet50 features, and (2048, 65, 65) for transmitting CCNet features. According to statistics, the values of $fr_1$, $fr_2$, and $fr_3$ are shown in Table \ref{tab:3}. 
\begin{table}[h]
\caption{Fire rate of different layers.}
\label{tab:3}
\centering
\resizebox{\linewidth}{15mm}{
\renewcommand\arraystretch{1.3}{
\begin{tabular}{c|c|c|c|c}
\hline
\hline
   Network &  Model  & $fr_1$   &  $fr_2$ & $fr_3$ \\
\hline
\hline
\multirow{2}*{\makecell[c]{ ResNet\\(2048,4,4)}}& SNN-SC (T = 4)& 0.25& 0.50 &0.23\\
                     \cline{2-5}
                    ~& SNN-SC (T = 8)& 0.27& 0.50 &0.24\\
                     \cline{2-5}
\hline

\multirow{2}*{\makecell[c]{ CCNet\\(2048,65,65)}}& SNN-SC (T = 4)& 0.22& 0.50 &0.21\\
                     \cline{2-5}
                    ~& SNN-SC (T = 8)& 0.25& 0.50 &0.24\\
                     \cline{2-5}
\hline
\end{tabular}
}}
\end{table}

Therefore, the parameters and computational complexity of SNN-SC and baseline models are shown in Table \ref{tab:2}. It can be observed that the number of parameters of SNN-SC is similar to that of CNN-Uni and CNN-NonUni, but lower than that of CNN-Bern. Moreover, despite the iterative nature of the SNN-SC, its computational complexity is still lower than that of the three baseline models due to the sparsity of the spiking features. In particular, when transmitting CCNet model features under low bandwidth conditions (T = 4), SNN-SC reduces the computation complexity by about 30\%, effectively reducing the model's energy consumption when it is deployed on the vehicle with limited computing power. Overall, using SNN-SC to transmit features over digital channels results in lower computational complexity than other models.
\begin{table}[h]
\normalsize
\caption{The complexity of the SNN-SC and Baselines.}
\label{tab:2}
\centering
\resizebox{\linewidth}{30mm}{
\renewcommand\arraystretch{1.3}{
\begin{tabular}{c|c|c|c}
\hline
\hline
   Network &  Model  & Param(M)   &  OP(GFLOPs) \\
\hline
\hline
\multirow{5}*{\makecell[c]{ResNet\\(2048,4,4)}}& CNN-Uni \cite{22}& 9.58& 0.31\\
                     \cline{2-4}
                    ~& CNN-NonUni \cite{92} & 9.58 & 0.31\\
                     \cline{2-4}
                    ~& CNN-Bern \cite{93} & 9.61 & 0.32 \\
                     \cline{2-4}
                    ~& SNN-SC (T = 4) & \textbf{9.58}& \textbf{0.23}\\
                     \cline{2-4}
                    ~& SNN-SC (T = 8) & \textbf{9.58}& \textbf{0.30}\\
                     \cline{2-4}
\hline

\multirow{5}*{\makecell[c]{CCNet\\(2048,65,65)}} & CNN-Uni \cite{22} & 9.58 & 80.99\\
                     \cline{2-4}
                    ~& CNN-NonUni \cite{92} & 9.58 & 80.99\\
                     \cline{2-4}
                    ~& CNN-Bern \cite{93} & 9.61 & 81.13\\
                     \cline{2-4}
                    ~& SNN-SC (T = 4) & \textbf{9.58} & \textbf{57.52}\\
                     \cline{2-4}
                    ~& SNN-SC (T = 8) & \textbf{9.58} & \textbf{75.16}\\
                     \cline{2-4}
\hline
\end{tabular}}
}
\end{table}

\section{Conclusion}
\label{sec:5}

In this paper, we propose an SNN-based semantic communication framework, SNN-SC, for transmitting features over digital channels. SNN-SC reduces the transmission overhead of AI models running on vehicle through CI and allows the cloud to maintain task performance in poor channel environments. By jointly optimizing the semantic encoder and the semantic decoder in a task-oriented manner, SNN-SC can extract compact semantic information effectively and transmit it robustly. Since the outputs of SNN are binary spikes, the SNN-SC framework can be directly applied to digital channels without explicit quantization. Furthermore, we propose a novel spiking neuron called IHF and a loss function based on maximizing entropy to improve the performance and robustness of SNN-SC. Both IHF and loss function were shown to have a positive effect on the model. Extensive experimental results show that SNN-SC outperforms the state-of-the-art digital semantic communication frameworks under poor channel conditions, and overcomes the `cliff effect' in the traditional separate coding method.
\bibliography{refs}
\bibliographystyle{IEEEtran}

\begin{IEEEbiography}[{\includegraphics[width=1in,height=1.25in,clip,keepaspectratio]{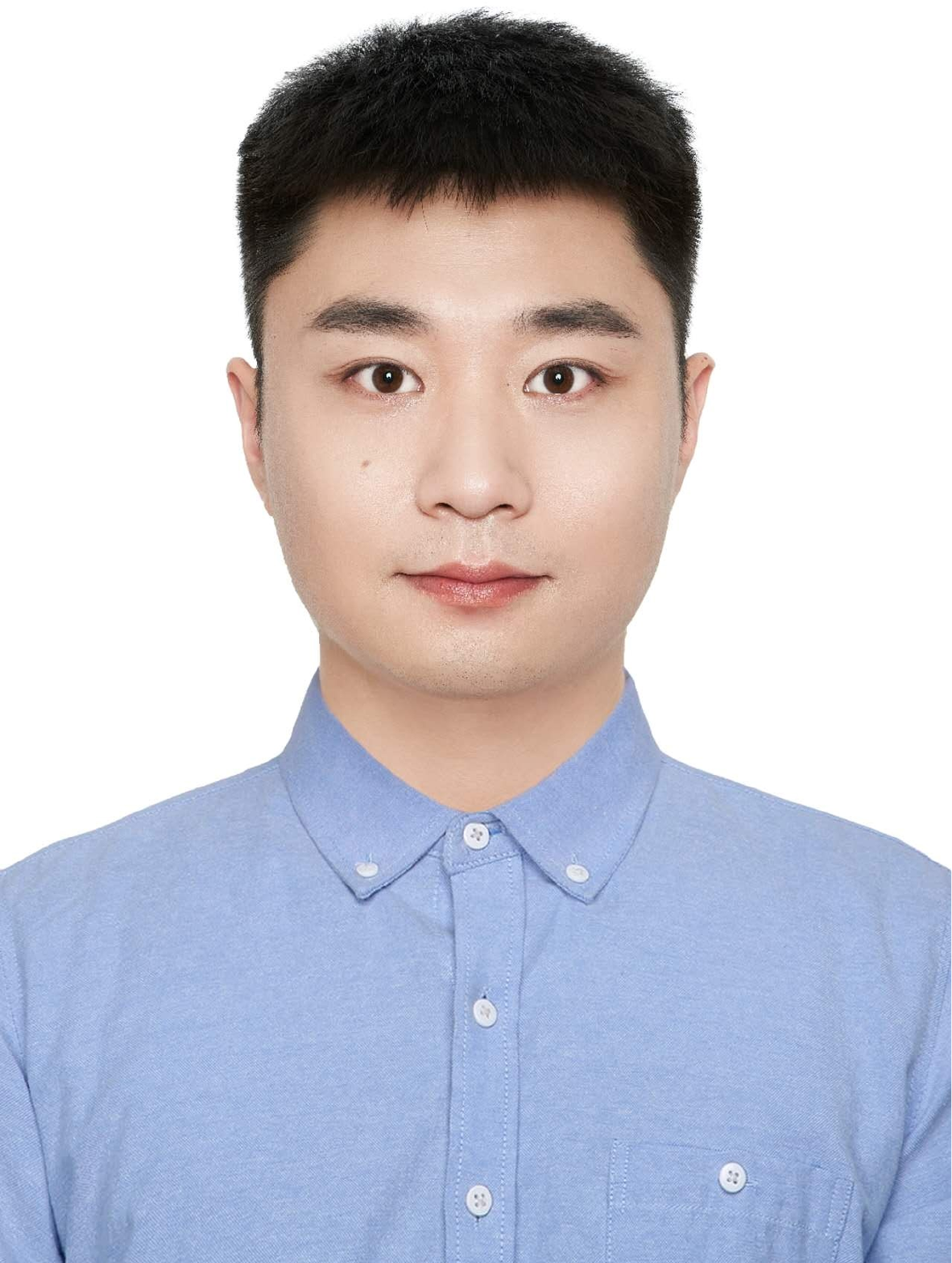}}]{Mengyang Wang}
received the B.E. degree from the School of Software, Dalian University of Technology, Dalian, China, in 2019. He is currently working toward the Ph.D. degree from the School of Computer Science, Harbin Institute of Technology (HIT), Harbin, China. His main research interests include joint source-channel coding and semantic communication.
\end{IEEEbiography}
\begin{IEEEbiography}[{\includegraphics[width=1in,height=1.25in,clip,keepaspectratio]{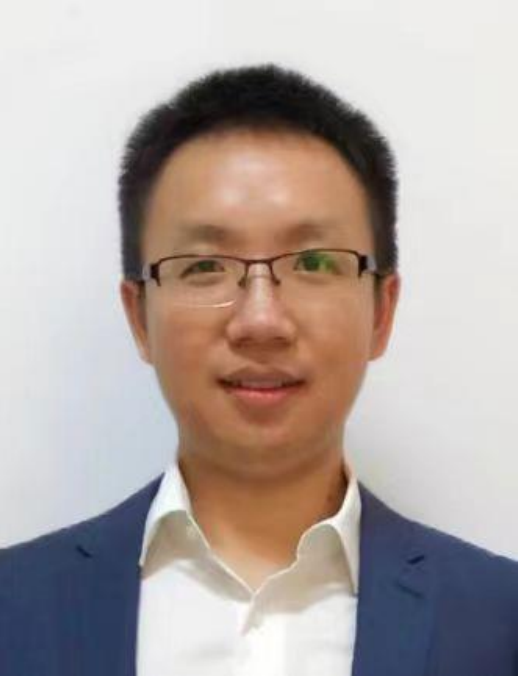}}]{Jiahui Li}
received the B.E. degree in electronic information science and technology and the Ph.D. degree in information and communication engineering from Tsinghua University, Beijing, China, in 2013, and 2018, respectively. He is currently a researcher at Huawei Technologies Co., Ltd. His research interests include joint source-channel coding and semantic communication.
\end{IEEEbiography}
\begin{IEEEbiography}[{\includegraphics[width=1in,height=1.25in,clip,keepaspectratio]{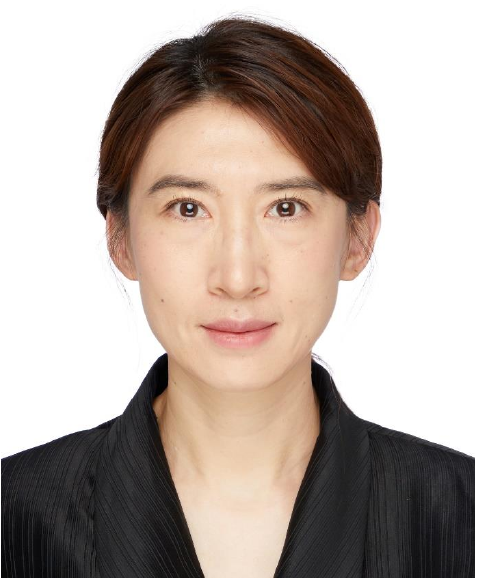}}]{Mengyao Ma}
received the Bachelor degree from the Department of Computer Science and Technology, Peking University, in 2003, and the Ph.D. degree from the Department of Computer Science and Engineering, Hong Kong University of Science and
Technology, in 2009. She is currently a technical expert at Huawei Technologies Co., Ltd. Her current research interests include joint source-channel coding and semantic communication.
\end{IEEEbiography}
\begin{IEEEbiography}[{\includegraphics[width=1in,height=1.25in,clip,keepaspectratio]{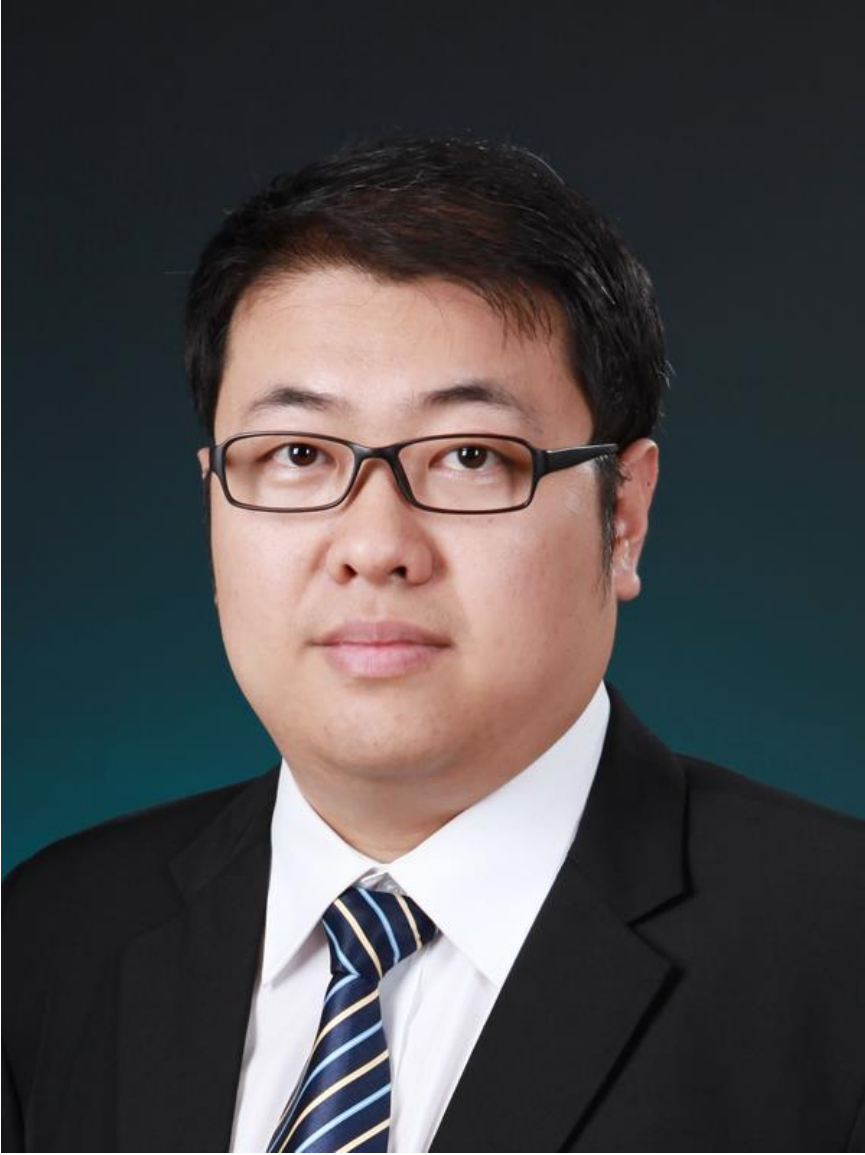}}]{Xiaopeng Fan}
(S’07–M’09–SM’17) received the B.S. and M.S. degrees from the Harbin Institute of Technology (HIT), Harbin, China, in 2001 and 2003, respectively, and the Ph.D. degree from The Hong Kong University of Science and Technology, Hong Kong, in 2009.

He joined HIT in 2009, where he is currently a Professor. From 2003 to 2005, he was with Intel Corporation, China, as a Software Engineer. From 2011 to 2012, he was with Microsoft Research Asia as a Visiting Researcher. From 2015 to 2016, he was with Hong Kong University of Science and Technology as a Research Assistant Professor. He has authored one book and more than 100 articles in refereed journals and conference proceedings. His current research interests
include video coding and transmission, image processing, and computer vision. He served as a Program Chair for PCM2017, Chair for IEEE SGC2015, and Co-Chair for MCSN2015. He was an Associate Editor of IEEE 1857 Standard in 2012. He received Outstanding Contributions to the Development of IEEE Standard 1857 by IEEE in 2013.
\end{IEEEbiography}

\end{document}